\documentclass[12pt]{iopart}
\usepackage{epsfig}
\usepackage{graphicx}

\begin{document}

\title{The Nearby, Young, Argus Association: Membership, Age, and Dusty Debris Disks}

\author{B. Zuckerman$^1$}

\address{$^1$Department of Physics and Astronomy, University of California, Los Angeles, CA 90095, USA}
\eads{\mailto{ben@astro.ucla.edu}}
\begin{abstract}
The reality of a field Argus Association has been doubted in some papers in the literature.  We apply Gaia DR2 data to stars previously suggested to be Argus members and conclude that a true association exists with age 40-50 Myr and containing many stars within 100 pc of Earth; $\beta$ Leo and 49 Cet are two especially interesting members.  Based on youth and proximity to Earth, Argus is one of the better nearby moving groups to target in direct imaging programs for dusty debris disks and young planets.
\end{abstract}
\maketitle

\section{Introduction}
The solar vicinity is blessed with an assortment of youthful stars with ages that span the range 10 to 200 Myr.   Within 100 pc of Earth a dozen or so coeval, co-moving groups were identified (Mamajek 2016) prior to the release of the Gaia DR2 catalog (Gaia Colaboration et al 2018; Lindegren et al 2018).   Many nearby stars exist that appear to be youthful but have not been placed into any of these groups. With the help of Gaia, new kinematic groups can be identified.   As recently as 20 years ago only a handful of youthful stars with reasonably reliable ages were known within 100 pc of Earth.  With Gaia and appropriate follow-up observations, it now appears likely that a few 1000 will ultimately be identified.

The present paper addresses one of the proposed youthful moving groups, first dubbed “Argus” by Torres et al (2008).   Their Argus Association was dominated by stars in the open cluster IC 2391 plus field stars $>$100 pc from Earth.  Subsequently, other astronomers have suggested additional Argus members; for example, Zuckerman et al (2011) and Zuckerman \& Song (2012) proposed more massive members closer to Earth than those listed by Torres et al.  One of these is $\beta$ Leo; with 20-20 hindsight, it might have been better to name the field association after $\beta$ Leo rather than Argus, similar, for example, to the $\beta$ Pictoris and AB Doradus moving groups, but it is now too late for that.  

The membership and even the reality of a field Argus Association (i.e. stars far from IC 2391) has been called into question by some astronomers.  For example, Bell et al (2015) state: “…it remains unclear whether this list represents a single, coeval population of stars, or even whether the association is in fact physical”.   Thus, Argus is not included in Mamajek’s (2016) catalog of stellar groups within 100 pc.  Nor is it included in the Gagne \& Faherty (2018) table of nearby young associations.   

We have gathered together a list of stars proposed in the literature to be Argus members.   We consider $\sim$40 of these to be true Argus members that are located within 118 pc of Earth (mean distance from Earth = 72.4 pc).  For a variety of reasons, we discard some previously proposed members.   It seems clear that a field Argus Association – with an age 40-50 Myr, comparable to the ages of the Tucana/Horologium and Columba Associations and much closer to Earth than IC 2391 -- really does exist.

\section{Sample selection}
The sample of proposed Argus members considered in this paper appear in one or more of:  Torres et al (2008), Riedel et al (2011), Zuckerman et al (2011), Shkolnik et al (2012), Zuckerman \& Song (2012), De Silva et al (2013), Moor et al (2013, 2016), and Elliot et al (2014, 2015, 2016).   We use radial velocity measurements given in these papers or in Vizier (many due to Gontcharov 2006) plus Gaia DR2 kinematic data to calculate Galactic space velocities UVW.   When available in the literature, we utilize age indicators -- such as the lithium 6707 $\AA$ absorption line and X-ray luminosity – to substantiate (or deny) membership in the Argus Association.   

A large list of proposed Argus Association members appears in Torres et al (2008).  The list is divided into members of the open cluster IC 2391 and field members.  A pre-Gaia distance to IC 2391 of 139 pc was assigned based on a proper motion convergence map.  The distance range of their proposed IC 2391 members is 120 to 154 pc, and of the proposed field members, 29 to 164 pc.   To distinguish field Argus stars from IC 2391 while at the same time retaining sufficient stars to construct a meaningful color-magnitude diagram, in the present paper we consider only proposed Argus members that are within 118 pc of Earth.  

\section{Results}
Stars considered in the present paper as potential Argus members appear in Table 1 and Figure 1.  Most of these stars are plotted in Figures 2 and 3, optical/infrared color-magnitude diagrams (CMDs).  The absolute Ks magnitudes (Ks$_{abs}$) and G-Ks colors for the plotted stars appear in Table 2.  All the proper motions and parallaxes used to calculate the UVWs given in Table 3 are from the Gaia DR2 catalog. 
($\beta$ Leo appears in Table 1 but not in Table 3 because $\beta$ Leo is not in Gaia DR2.)  The radial velocities listed in Table 3 are an error-weighted combination of values given in the references listed in Section 2 plus those given in Vizier.  Table 3 contains UVW values for 37 stars of which 31 plus $\beta$ Leo are included in a calculation of the mean UVW of the field Argus stars; reasons for non-inclusion of seven stars are given in the Notes to Table 3.

As given in the footnote to Table 3, the mean UVW for the 32 field stars within 118 pc of Earth is -22.5+/-1.2, -14.6+/-2.1, -5.0+/-1.6.  This compares well with the UVW of the Torres et al (2008) Argus Association:  -22.0+/-0.3, -14.4+/-1.3, -5.0+/-1.3.  Two thirds of the total of 64 stars listed in their Argus Association are at distances greater than 118 pc; 35 in IC 2391 and 8 in the field.  Obviously, the agreement in UVW over the entire range of distances considered is excellent.

The highest concentration of Argus stars is located near IC 2391 (8h40m,-53d; see Figure 1).  We note an interesting coincidence.   The 200 Myr old Carina-Near Moving Group (Zuckerman et al 2006), with nucleus located near 7h40m R.A. and -55 deg declination, has a similar UVW to Argus, but is substantially closer ($\sim$30 pc from Earth) and older.   The mean UVW of Carina-Near is approximately -26, -18, -2 km/s.   Thus, identification of young stars in the general vicinity of 8h and -50 deg should take care to distinguish between the two moving groups.

Measurements of youth indicators are available in the literature for most proposed Argus members; two indicators – lithium absorption and X-ray luminosity -- are listed in Table 4. Others, such as excess infrared emission and H$\alpha$ emission, are given in the Comments column in Table 1 and/or in the Appendix.  For lithium, we compare the equivalent width (EW) of the 6707 $\AA$ line with the EW presented in Figure 3 of Zuckerman \& Song (2004) for stars of comparable spectral type that are located in various known moving groups.   For X-ray luminosity (L$_{x}$/L$_{bol}$), we compared the stars in Table 4 with those in Figure 4 of Zuckerman \& Song (2004).  The Li measurements indicate that the Table 4 stars are significantly younger than the Pleiades and comparable in age to stars in Tuc/Hor. 

Most of the putative Argus stars lie near the plotted 40 Myr isochrone of Tognelli et al (2011, Figure 2).  While the Baraffe et al (2015, Figure 3) evolutionary models also fit most of the intermediate mass stars, the theoretical isochrones dip down too quickly for low mass stars.  Clearly, the intermediate mass stars present a range of luminosities at a given G-Ks color.  However, as discussed in Section 4, plausible physical processes can potentially account for this luminosity spread.

\section{Discussion}
The primary goal of the present paper is to update and organize information available in the published literature so as to clarify whether Argus is a real association of young stars and to establish which stars are likely to be members.  The information presented in the various figures and tables and in the discussion below, establish, in the opinion of the author, the reality of a young nearby moving group.   Likely members are listed in Appendix A.   Those stars that have been suggested  to be members but now appear, for one reason or another to be unlikely Argus members, are listed in Appendix B.   As initially pointed out by Torres et al (2008) the Argus field stars are very likely associated with and coeval with the open cluster IC 2391. 

Argus membership is based on UVW, various youth indicators (e.g., lithium abundance), and location on  CMDs.   Section 3 includes a discussion of UVW and youth indicators.  We now consider the CMD displayed in Figures 2 and 3.   In general, for youthful moving groups, there is some spread in absolute magnitude at a given color.  Thus, the fact that all likely members of Argus do not lie along a single isochrone is not unusual.  Unrecognized binarity is one reason why stars can lie above the single star isochrone.

There are good reasons to believe that  even CD-43 3604, CD-57 2315, and CPD-62 1197 -- three K-type stars that lie near the 10 Myr isochrone in Figures 2 and 3 -- are actually members of Argus and thus have ages near 40 Myr.   One such reason involves Occam's razor.  These stars all lie in the heart of the Argus sky plane distribution (Figure 1) and they have UVW consistent with Argus membership. The three stars have distances from Earth of 85, 94 and 110 pc, respectively.  Over the entire sky, what 10 Myr old stars are known within 100 pc of Earth?  Only members of TW Hya and $\eta$ Cha.  The UVW of the three stars totally disagree with the UVW of TW Hya and $\eta$ Cha.  So if these stars are not members of Argus then we would have the most peculiar situation of uniquely young and nearby stars not in TW Hya or $\eta$ Cha but with UVW that just happen to agree with Argus.  Thus, Occam’s razor strongly points to Argus membership even if it is unclear why these three stars plot so high on the CMD.  

Table 5 suggests one possible contributor to the high location on the CMD of some stars with G-Ks between about 1.5 and 2.5.  Stars in this range of G-Ks have spectral types between late-G and mid-K.  Table 5 compares vsini and Li 6707$\AA$ EW values for stars that fall near the 40 Myr isochrone with those that lie well above it.  

For five stars that plot high on Figures 2 and 3, the mean vsini = 52.3+/-30.0 km/s, while for eight stars that appear lower (near the 40 Myr isochrone), the mean vsini = 18.4+/-14.8 km/s.  The plus/minus are the standard deviations; the uncertainty in the mean velocities will of course be much less. The more luminous stars are, on average, rotating much more rapidly.   

Comparison of lithium 6707 $\AA$ EW between stars with G-Ks in the range 1.5 to 2.5 that lie high and low on Figures 2 and 3, suggests that, on average, the EW is larger for the more luminous (larger) atars.  Specifically, for six stars thst are high on the color-mgnitude diagram, the mean Li EW equals 279+/-31 m$\AA$, while for ten stars that lie nearer to the 40 Myr isochrone, the mean EW equals 223+/-38 m$\AA$, where again the plus/minus are standard deviations.  Although not as striking as the differences in vsini, these Li results indicate that, on average, more rapidly rotating stars burn lithium more slowly.  

Similar relationships among rotation rate, luminosity, and lithium abundances have been demonstrated for stars in the Pleiades (Somers \& Stassun 2017); they suggest a magnetic origin for the observed patterns.
Although an explanation for the patterns indicated in Table 5 is beyond the scope of the present work, w mention a physical mechanism suggested by Joel Kastner (2018, private communication) that might plausibly account for the various variations.   Strong magnetic fields in young stars might supply additional pressure that suppresses convection that would otherwise cause lithium to be burned while, at the same time, providing extra pressure that inflates the radii of such stars.

The presence of excess infrared emission due to orbiting dust grains is correlated with stellar age (e.g., Table 11 in Zuckerman et al 2011).  Most of the A-type stars in Argus display excess IR emission; these excesses are noted in Table 1 and in Appendix A.   After excluding HD 188728 and 192640 -- that may be older than Argus stars (see Appendix B) -- of the 10 A-type stars that we designate as probable members of Argus, 8 or 80\% have excess IR emission.  This is the highest percentage of dusty A-type stars of all of the nearby young moving groups, and firmly attests to the youth of the Argus stars.  Again one may appeal to Occam’s razor: That 80\% of the 10 A-stars have excess IR emission and UVW in reasonable agreement with Argus indicates a young coeval population.  In addition, the solar-type stars HD 61005 and HD 84075 display excess IR emission. 

De Silva et al (2013), their Table 3, derive a ``convergence age'' for the field Argus stars and for IC 2391.  They utilize a “kinematical convergence” method first introduced by Torres et al (2006) to estimate distances to Argus members and thence an age.  This was of course before parallaxes from Gaia were available.  With this method an age of 26 Myr for both Argus field stars and IC 2391 was deduced (Table 3 in De Silva et al).

Other ages for IC 2391 have been suggested in the literature; those of Barrado et al (2004; 50+/-5 Myr) and Randich et al (2018; 44+/-5 Myr) appear to be representative.  Based on the above discussion, and the Figures and Tables in the present paper, the age of the much closer Argus field stars appears to be comparable to that of IC 2391.   The Gaia DR2 catalog, when fully exploited, should enable comparison of color-magnitude diagrams for the major nearby young moving groups, including Argus.  These should firmly establish the relative ages of these groups.



\section{Conclusions}

The Argus Association was first proposed by Torres et al (2008) to be a mixture of the open cluster IC 2391 and a group of field stars with similar Galactic space motions UVW, most of which were closer to Earth than is IC 2391 (located $\sim$140 pc away).  Membership in, and even the reality of a coeval Argus field association has been somewhat uncertain.  In the present paper we have used the Gaia DR2 catalog and supplemental information from the published literature to develop a list of field stars that comprise a reliable membership base.  The primary criteria for group membership are location on a Gaia-based optical/infrared color-magnitude diagram, lithium abundance, and UVW. A group of 40 or so likely Argus field stars within 120 pc of Earth has a mean distance from Earth of  72.4 pc and a likely age somewhere between 40 and 50 Myr. While the (model dependant)  absolute age is somewhat uncertain, the age of Argus relative to other young moving groups, such as $\beta$ Pictoris and Tucana/Horologium, can be established quite reliably from comparison of  lithium abundances and Gaia-based color-magnitude diagrams.   

One noteworthy feature of the proposed Argus Association is the high percentage -- perhaps as much as 80\% -- of A-type members with excess infrared emission due to orbiting dust particles.   This attests to the youth of these stars.

We thank Carl Melis for kindly obtaining a high resolution spectrum of 
TYC 155-2167-1, Joel Kastner, Navya Nagananda, and Germano Sacco for their help with Figures 2 and 3 and other assistance, Inseok Song for help with Tables 1 and 4 and Figure 1, Jonathan Gagne for various assistance, and the referee for suggestions that helped to improve the paper.  This research was supported in part by grants to UCLA from NASA. 

\appendix



\section*{References}
\begin{harvard}

\item[Baraffe, I., Homeier, D., Allard, F. \& Chabrier, G. 2015, A\&A 577, A42]
\item[Barrado y Navascués, D,, Stauffer, J. R. \& Jayawardhana, R. 2004, ApJ 614, 386]
\item[Bell, C. P. M., Mamajek, E. E. \& Naylor, T. 2015, MNRAS 454, 593]
\item[da Silva, L.,Torres, C. A. O., de La Reza, R. et al 2009 A\&A 508, 833]
\item[De Silva, G. M., D'Orazi, V., Melo, C. et al 2013, MNRAS 431, 1005]
\item[Elliott, P., Bayo, A., Melo, C. et al 2014, A\&A 568, A26]
\item[Elliott, P., Bayo, A., Melo, C. et al 2016, A\&A 590, A13]
\item[Elliott, P., Huélamo, N., Bouy, H. et al 2015, A\&A 580, A88]
\item[Gagne, J. \& Faherty, J. 2018, ApJ 862, 138]
\item[Gaia Collaboration; Brown, A. G. A., Vallenari, A., Prusti, T., et al. 2018, arXiv:1804.09365]
\item[Gontcharov, G. 2006, Astronomy Letters 32, 759]
\item[Kiraga, M.2012, Acta Astronomica 62, 67]
\item[Lindegren, L., Hernandez, J., Bombrun, A. et al. 2018, arXiv:1804.09366]
\item[Mamajek, E. E. 2016, in IAU Symp. 314, Young Stars $\&$ Planets Near the Sun, ed. J. H.] Kastner et al (Cambridge: Cambridge Univ. Press), 21
\item[Moór, A., Kóspál, Á., Ábrahám, P. et al 2016, ApJ 826, 123]
\item[Moór, A., Szabó, G. M., Kiss, L. L. et al 2013, MNRAS 435, 1376]
\item[Neugebauer, G. \& Leighton, R. 1969, Two-Micron Sky Survey, NASA SP-3047]
\item[Randich, S., Tognelli, E., Jackson, R. et al 2018, A\&A 612, A99]
\item[Riaz, B., Gizis, J. E. \& Harvin, J. 2006, AJ 132, 866]
\item[Riedel, A. R., Murphy, S. J., Henry, T. et al 2011, AJ 142, 104]
\item[Shkolnik, E. L., Anglada-Escudé, G., Liu, M. C. et al, 2012, ApJ 758, 56]
\item[Somers, G. \& Stassun, K. 2017, AJ 153, 101]
\item[Tognelli, E., Prada Moroni, P. G., Degl'Innocenti, S. 2011, A\&A 533, A109]
\item[Torres, C. A. O., Quast, G. R., da Silva, L. et al 2006, A\&A 460, 695]
\item[Torres, C. A. O., Quast, G. R., Melo, C. H. F, \& Sterzik, M. F. 2008, Handbook of Star Forming] Regions, Volume II: The Southern Sky ASP Monograph Publications, Ed. B. Reipurth, p.757
\item[White, R. J., Gabor, J. M. \& Hillenbrand, L. A. 2007, AJ 133, 2524]
\item[Zuckerman, B., Bessell, M. S., Song, I., Kim, S. 2006, ApJ 649, L115]
\item[Zuckerman, B., Forveille, T. \& Kastner, J. H. 1995, Nature 373, 494]
\item[Zuckerman, B., Rhee, J. H., Song, I. \& Bessell, M. S. 2011, ApJ 732, 61]
\item[Zuckerman, B. \& Song, I. 2004, ARA\&A 42, 685]
\item[Zuckerman, B. \& Song, I. 2012, ApJ 758, 77]

\end{harvard}

\clearpage

\begin{figure}
\includegraphics[width=150mm]{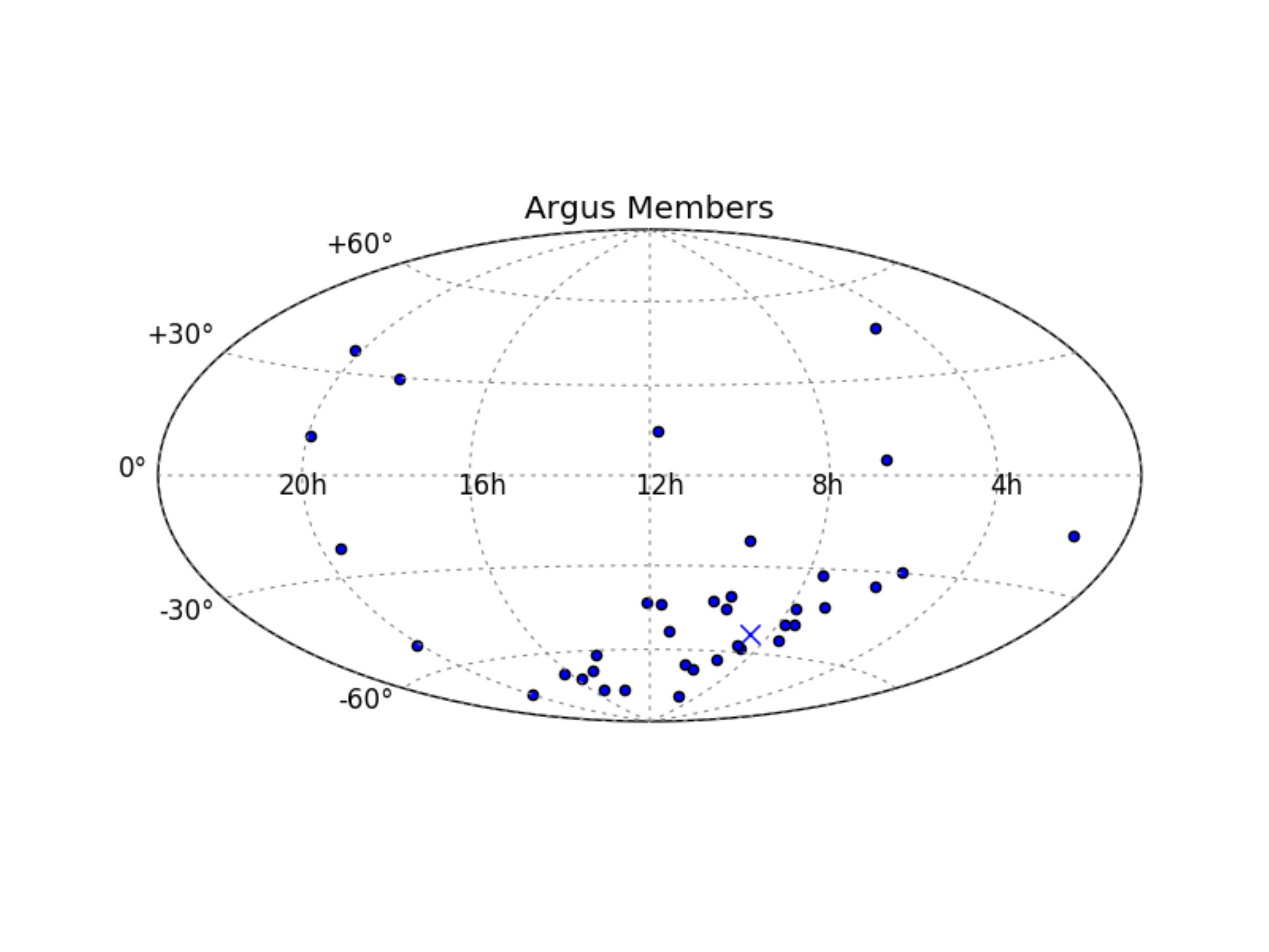}

\caption{\label{figure 1} Positions of the candidate Argus members listed in Table 1, shown here in Aitoff projection.  The position of open cluster IC 2391 is indicated with a cross.  Of the six stars with positive declinations, two (HD 188728 and HD 192640) are doubtful members (see Appendix B).  Another star at positive declination, $\beta$ Leo, is only 11 pc from Earth so that its sky plane position could easily differ from that of members that are at substantially greater distances.}
\end{figure}
\clearpage

\begin{figure}
\includegraphics[width=150mm]{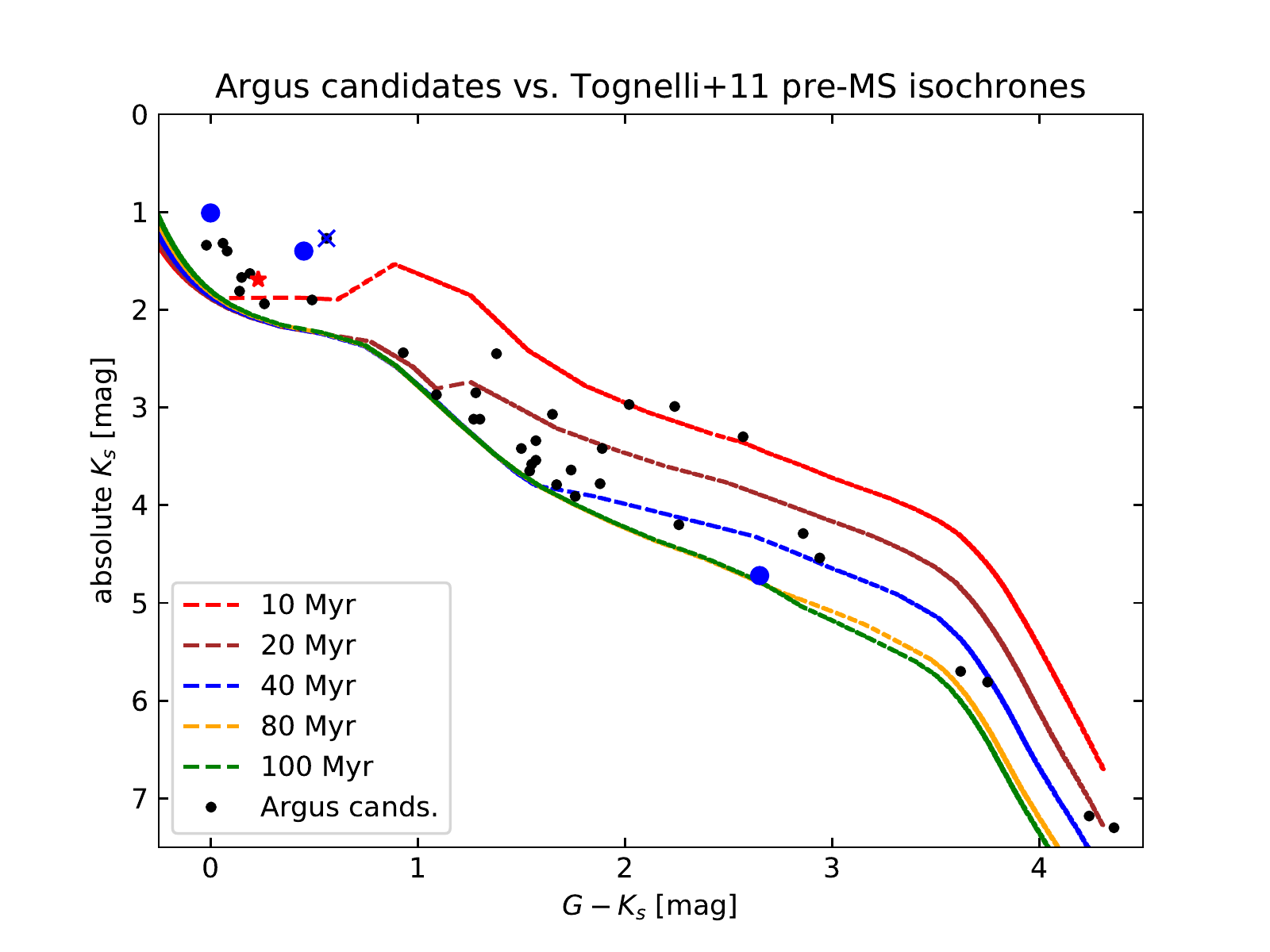}


\caption{\label{figure 2} Color-magnitude diagram based on data in Table 2.  The isochrones from top to bottom are for 10, 20, 40, 80 and 100 Myr,
respectively, and are for solar metallicity; they
are from Tognelli et al. (2011).  The red star plotted at Ks$_{abs}$ = 1.69 and G-Ks = 0.23 is $\beta$ Leo (= HD 102647); the reason for the special symbol is given in Appendix A. The star at Ks$_{abs}$ = 1.27 and G-Ks = 0.56 marked with an X is HD 168913 which is composed of two nearly equal luminosity late A-type stars.  When corrected downward by 0.7 magnitudes, the location of HD 168913 on the plot is consistent with those of other A-type members of Argus. The A-type stars HD 188728 at Ks$_{abs}$ = 1.01 and G-Ks = 0, and HD 192640 at Ks$_{abs}$ = 1.4 and G-Ks = 0.45 marked with large blue circles, also sit well above the other A-stars on the plot.  Since neither of these stars is known to be a close binary, they may be older than the Argus stars and thus not members of the Association, as noted in Appendix B.  The star at Ks$_{abs}$ = 4.72 and G-Ks = 2.65 marked with a large blue circle is CD-52 9381.   Because of its low position on the CMD it may be older than Argus members (see discusion in Appendix B).   The three stars with G-Ks between 2 and 2.6 that sit on the 10 Myr isochrone are very likely actually 40 Myr old members of Argus (see discusion in Section 4).} 
\end{figure}
\clearpage

\begin{figure}
\includegraphics[width=150mm]{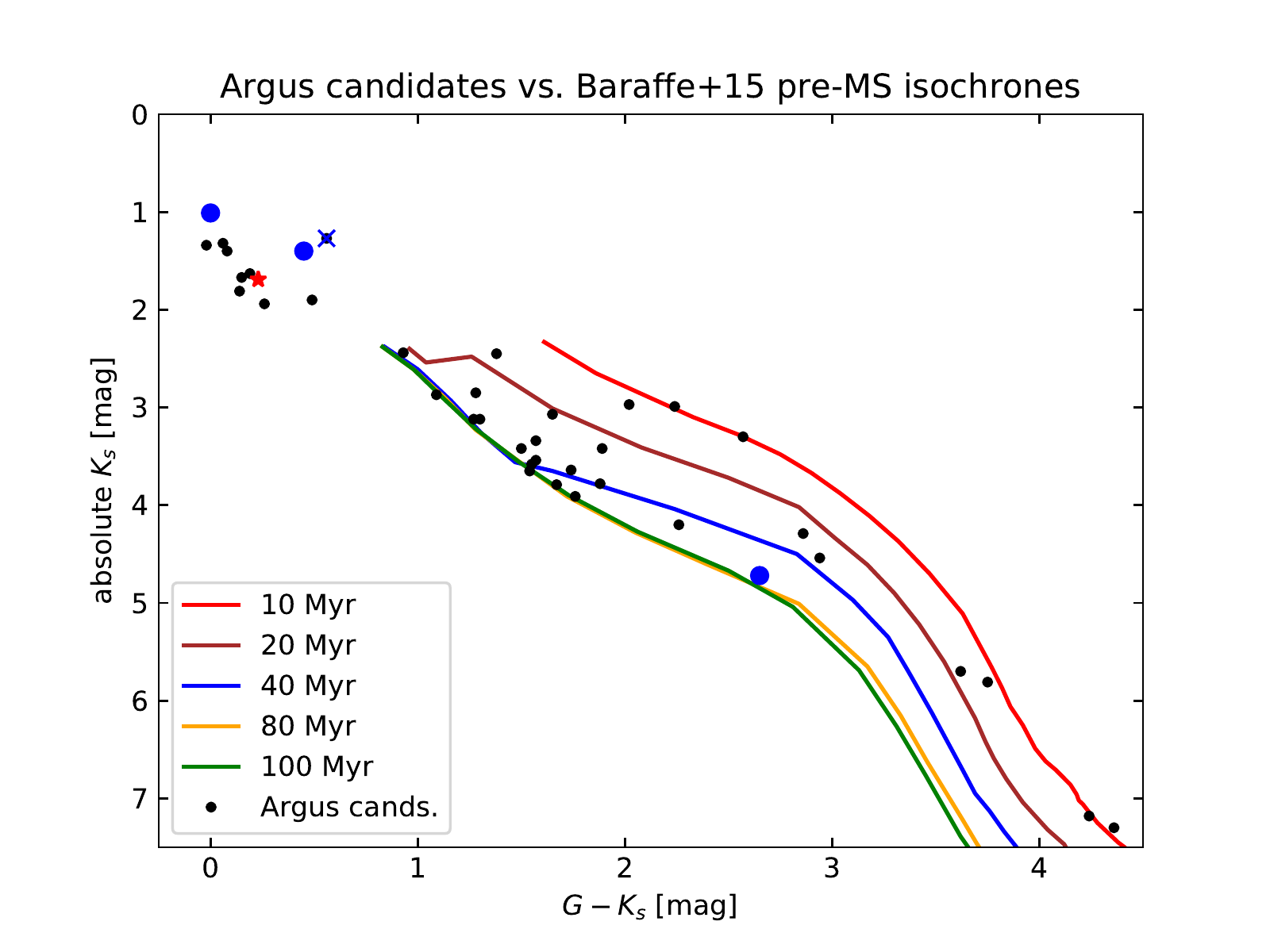}


\caption{\label{figure 3}  Color-magnitude dagram based on data in Table 2.  The isochrones
are from Baraffe et al. 2015. See also caption to Figure 1.}
\end{figure}

\clearpage

\begin{table}
\caption{Suggested Members of the Argus Association}
\begin{tabular}{@{}lcccc}
\br
Star & Spectral Type & Right Ascension & Declination & Comments  \\

\mr
49 Cet & A1 & 01:34:37.8 & -15:40:34.9 & IR emission \\
HD 32296 & A2 & 05:04:16.2 & +45:46:45.6 & IR emission \\
CD-29 2360 & K3Ve & 05:34:59.2 & -29:54:04.0 &  \\
AP Col & M5 &  06:04:52.1 &  -34:33:35.8  &  \\
TYC 155-2167-1 & late-K &  06:41:09.2 & +04:47:18.7 &  \\
CD-42 2906 & K1 & 07:01:53.4 & -42:27:56.2 & \\
CD-42 2906B  & early-M & 07:01:53.9 &  -42:27:58.3   & \\
CD-48 2972 & G8  & 07:28:22.0 &  -49:08:37.6 & \\
HD 59659 & F7 & 07:28:30.1 & -49:08:58.9 & \\
CPD-54 1295 & late-G &  07 31 19.9 & -54:47:02.8 & 3.4'' binary \\
HD 61005 & G8 & 07:35:47.5 & -32:12:14.0 & IR emission \\
CD-48 3199 & G7  & 07:47:26.0 & -49:02:51.1 &  \\
CD-43 3604 & K4Ve  & 07:48:49.8 & -43:27:05.6 & \\
CD-58 2194 & G5 & 08:39:11.5 & -58:34:28.0 & \\
CD-57 2315 & K3Ve & 08:50:08.1 & -57:45:59.2 & \\
CPD-62 1197 & K0Ve  &  09:13:30.3 & -62:59:09.4 & \\
HD 84075 & G2 & 09:36:17.8 & -78:20:41.6 & IR emission \\
BD-20 2977 & G9V &  09:39:51.4 & -21:34:17.4 & \\
CD-39 5833 & K0V & 09:47:19.9 & -40:03:09.8 & \\
HD 85151 & F8/G0 & 09:48:43.3 & -44:54:08.4 & 2'' binary \\
HD 309851 & G1V &  09:55:58.3 & -67:21:21.8 & \\
HD 88955 & A2 & 10:14:44.2 & -42:07:19.0 & IR emission \\
HD 90874 & A2 & 10:27:25.3 & -65:42:16.8 & IR emission \\
TYC 8617-909-1 & M1 & 11:23:47.0 & -52:57:39.4 & \\
HD 101615 & A0 & 11:41:19.8 & -43:05:44.4 & \\
$\beta$ Leo &  A2.5 & 11:49:03.6 & +14:34:19.4 & IR emission \\
CD-42 7422 & K0V &  12:06:32.9 & -42:47:50.8 &  \\
CD-75 652 &  G1V & 13:49:12.9 & -75:49:47.4 & \\
CD-75 652B & mid-M  &  13:49:07.3 &  -75:49:52.3 & \\
HD 123058 & F4 & 14:07:29.3 & -61:33:44.2 & \\
HD 129496 & F7 & 14:46:21.4 & -67:46:15.6 & SB1? \\
NY Aps & G9V & 15:12:23.4 & -75:15:15.6 & \\
HD 138965 & A1 & 15:40:11.6 & -70:13:40.4 & IR emission \\
HD 145689 & A4 & 16:17:05.4 & -67:56:28.6 & IR emission \\
HD 168913 & A7+A9 & 18:20:57.0 & +29:51:32.1 & SB2 \\
2MASSJ1931-2134 & M2.4 & 19:31:24.3 &  -21:34:22.8 & \\
HD 188728 & A1 & 19:56:14.3 & +11:25:25.4 & member? \\
HD 188228 & A0 & 20:00:35.6 & -72:54:37.8 & IR emission \\ 
CD-52 9381 & K6Ve & 20:07:23.8 & -51:47:27.3 & member? \\
HD 192640 & A2 &  20:14:32.0 & +36:48:22.7 & member? \\

\br
\end{tabular}
\end{table}

\clearpage

\noindent Notes to Table 1 $-$ The stellar positions are equinox 2000.  HD 188728, CD-52 9381, and HD 192640 may not be Argus members (see Appendix B).  We consider all other stars in this table to more likely be Argus members than not.  However, in a few cases membership appears to be marginal (see Appendix A).

\clearpage

\begin{table}
\caption{Infrared/Optical Colors}
\begin{tabular}{@{}lcccc}
\br
Star & Ks & Ks$_{abs}$ & G &  G-Ks  \\
\mr
49 Cet & 5.45 & 1.67 & 5.60 & 0.15 \\
HD 32296 & 5.98 & 1.90 & 6.47 & 0.49 \\
CD-29 2360 & 7.99 & 4.20 & 10.25 & 2.26 \\
AP Col & 6.87 & 7.18 & 11.11 & 4.24 \\
TYC 155-2167-1 & 8.69 & 4.29 & 11.55 & 2.86 \\
CD-42 2906 & 8.65 & 3.91 & 10.41 & 1.76 \\
CD-42 2906B & 10.55 & 5.81 & 14.30 & 3.75 \\
CD-48 2972 & 8.06 & 3.34 & 9.63 & 1.57 \\
HD 59659 & 7.58 & 2.87 & 8.67 & 1.09 \\
CPD-54 1295 & 8.75 & 3.58 & 10.30 & 1.55 \\
HD 61005 & 6.46 & 3.65 & 8.00 & 1.54 \\
CD-48 3199 & 8.84 & 3.54 & 10.41 & 1.57 \\
CD-43 3604 & 8.19 & 3.30 & 10.76 & 2.57 \\
CD-58 2194 & 8.37 & 3.07 & 10.02 & 1.65 \\
CD-57 2315 & 7.96 & 2.99 & 10.20 & 2.24 \\
CPD-62 1197 & 8.32 & 2.97 & 10.34 & 2.02 \\
HD 84075 & 7.16 & 3.12  & 8.43 & 1.27 \\
BD-20 2977 & 8.38 & 3.64 & 10.12 & 1.74 \\
CD-39 5833 & 8.88 & 3.79 & 10.55 & 1.67 \\
HD 85151 & 7.58 & 2.45 & 9.44 & 1.86 \\
HD 309851 & 8.31 & 3.12 & 9.61 & 1.30 \\
HD 88955 & 3.77 & 1.34 & 3.75 & -0.02 \\
HD 90874 & 5.79 & 1.63 & 5.98 & 0.19 \\
TYC 8617-909-1 & 7.95 & 4.54 & 10.89 & 2.94 \\
HD 101615 & 5.44 & 1.40 & 5.52 & 0.08 \\
CD-42 7422 & 8.58 & 3.42 & 10.47 & 1.89 \\
CD-75 652 & 8.02 & 3.42 & 9.52 & 1.50 \\
CD-75 652B  & 11.90 & 7.30 & 16.26 & 4.36 \\
HD 123058 & 6.72 & 2.44 & 7.65 & 0.93 \\
HD 129496 & 7.43 & 2.85 & 8.71 & 1.28 \\
NY Aps & 7.38 & 3.78 & 9.26 & 1.88 \\
HD 138965 & 6.27 & 1.81 & 6.41 & 0.14 \\
HD 145689 & 5.66 & 1.94 & 5.92 & 0.26 \\
HD 168913 & 4.99 & 1.27 & 5.55 & 0.56 \\ 
2MASSJ1931-2134 & 7.83 & 5.70 & 11.45 & 3.62 \\
HD 188728 & 5,25 & 1.01 & 5.25 & 0 \\
HD 188228 & 3.89 & 1.32 & 3.86 & 0.06 \\
CD-52 9381 & 7.39 & 4.72 & 10.04 & 2.65 \\
HD 192640 & 4.42 & 1.4 & 4.87 & 0.45 \\

\br
\end{tabular}
\end{table}

\clearpage

\noindent Notes to Table 2 $-$ The Ks values are in magnitudes from the 2MASS catalog.  The 
G values are in magnitudes from the Gaia DR2 catalog.   For $\beta$ Leo (= HD 102647), see discussion in Appendix A.

\clearpage

\begin{table}
\caption{Galactic Space Velocities UVW}
\begin{tabular}{@{}lccc}
\br
Star & Radial  & UVW  & UVW uncertainty \\
& Velocity (km/s)  & km/s  & km/s    \\
\mr
49 Cet & 10.3+/-0.7 & -21.9,-15.6,-5.6 & 0.2, 0.1, 0.7 \\ 
HD 32296 & 19.4+/-1.0 & -24.8,-12.1,-5.9 & 0.9, 0.3, 0.1 \\
CD-29 2360 & 25.8+/-0.3 & -21.3,-17.5,-4.9  & 0.2, 0.2, 0.1 \\
AP Col & 22.4+/-0.3 & -22.4,-13.5,-4.1 & 0.1, 0.2, 0.1  \\
TYC 155-2167-1 & 27.8+/-1.6 &  -24.0,-14.2,-4.2 & 1.4, 0.7, 0 \\
CD-42 2906 & 23.6+/-0.2 & -21.4,-17.6,-5.3 & 0.1, 0.2, 0.1 \\
CD-48 2972 & 18.8+/-0.5 & -23.7,-14.5,-5.7 & 0.1, 0.5, 0.1 \\
HD 59659 & 18.8+/-0.5 & -24.0,-14.5,-5.6 & 0.1, 0.5, 0.1 \\
CPD-54 1295 & 19.8+/-0.4 & -20.8,-17.9,-5.5 & 0.1, 0.4, 0.1 \\
HD 61005 & 22.4+/-0.2 & -23.3,-14.0,-4.1 & 0.1, 0.2, 0 \\
CD-48 3199 & 17.6+/-0.3 & -23.9,-13.8,-5.8  & 0.1, 0.3, 0.1 \\
CD-43 3604 & 18.0+/-3.0 & -23.0,-13.3,-2.2  & 0.7, 2.9, 0.5 \\
CD-58 2194 &  14.0+/-1.5 & -23.0,-14.7,-8.7 & 0.1, 1.5, 0.3 \\
CD-57 2315 & 11.9+/-0.4 & -22.2,-13.5,-4.3 & 0.4, 0.4, 0.2 \\
CPD-62 1197 & 12.7+/-2.0 & -23.5,-16.5,-7.2 & 0.4, 1.9, 0.3 \\
HD 84075 & 5.2+/-0.1 & -22.8,-13.7,-6.4 & 0.1, 0.1, 0 \\
BD-20 2977 & 18.3+/-0.6 & -21.5,-16.7,-4.2 & 0.2, 0.5, 0.2 \\
CD-39 5833 & 14.7+/-0.3 & -22.6,-15.7,-5.4 & 0.1, 0.3, 0.1 \\
HD 85151 & 14.3+/-0.5 & -22.8,-16.0,-4.0 & 0.1, 0.5, 0.1 \\
HD 309851 & 7.4+/-0.4 & -23.9,-14.3,-5.9 & 0.1, 0.4, 0.1 \\
HD 88955 & 7.7+/-2.7 & -21.3,-10.7,-4.8 & 0.3, 2.6, 0.6 \\
HD 90874 & 7.1+/-0.6 & -22.5,-14.2,-8.7 & 0.2, 0.6, 0.1 \\
TYC 8617-909-1 & 3.9+/-0.5 & -22.8,-13.1,-3.8 & 0.2, 0.5, 0.1 \\
HD 101615 & 9.0+/-3.0 & -19.2,-18.0,-2.5 & 1.0, 2.7, 0.9 \\
CD-42 7422 & 1.0+/-1.5 & -23.7,-14.3,-5.7 & 0.6, 1.3, 0.5 \\
CD-75 652 & -1.0+/-0.2 & -22.1,-13.5,-5.7 & 0.1, 0.2, 0.1 \\
HD 123058 & -4.9+/-0.4 & -22.1,-13.4,-2.7 & 0.3, 0.3, 0 \\
HD 129496 & -8.0+/-4.0 & -24.0,-11.1,-3.7 & 2.7, 2.9, 0.5 \\
NY Aps & -3.7+/-0.3 & -22.4,-13.3,-4.5 & 0.2, 0.2, 0.1 \\
HD 138965 & -2.0+/-4.3 & -19.3,-15.5,-6.1 & 3.1, 2.9, 0.9 \\
HD 145689 & -9.0+/-4.3 & -23.4,-12.3,-4.6 & 3.2, 2.7, 1.0 \\
HD 168913 & -21.9+/-3.0 & -23.1,-11.7,-2.5 & 1.5, 2.4, 1.0 \\
2MASSJ1931-2134 & -25.6+/-1.5 & -23.5, -17.9, -2.5 & 1.4, 0.4, 0.5 \\
HD 188728 & -28.0+/-4.2 & -23.9, -17.9, -3.8 & 2.6, 3.2, 0.7 \\
HD 188228 & -6.7+/-0.7 & -21.1,-10.3,-4.1 & 0.5, 0.4, 0.4 \\
CD-52 9381 & -13.5+/-0.7 & -24.3,-17.2,-5.7 & 0.6, 0.1, 0.4 \\
HD 192640 & -17.3+/-2.8 & -22.3, -11.8, -3.5 & 0.7, 2.7, 0.1 \\

\br
\end{tabular}
\end{table}

\clearpage

\noindent Notes to Table 3 $-$ HD 188728, CD-52 9381, and HD 192640 may not be Argus members.  See discussion in Appendix B.  None of these stars are included in the calculation of the mean UVW for the Argus field stars within 118 pc of Earth.  Also not included in the calculation of the mean are HD 129496, HD 138965, and HD 145689 because of their relatively large uncertainties in UVW.   The mean UVW for the remaining 31 stars plus HD 102647 ($\beta$ Leo; see Appendix A) is -22.52+/-1.2, -14.57+/-2.1, -5.01+/-1.6.  To calculate mean distance from Earth (= 72.4 pc) we used 33 stars from Table 3 plus $\beta$ Leo.  HD 188728, CD-52 9381, and HD 192640 were excluded because they may not be Argus members and CD-48 2972 was excluded because it is a companion of HD 59659 (which is included).

\clearpage

\begin{table}
\caption{Lithium Equivant Width \& X-ray Luminosity}
\begin{tabular}{@{}lcccc}
\br
Star & Li 6707 $\AA$ & Lithium  & log [L$_{x}$/L$_{bol}$] &  X-Ray   \\
&  EW (m$\AA$) & Reference   &  &  Reference \\
\mr
CD-29 2360 &  180 & 1 & -3.2 & 2 \\
AP Col & 280 & 3 & -2.95 & 3,4 \\
TYC 155-2167-1 & 235 & 5 & -3.16 & 10 \\
CD-42 2906 & 275 & 1 & -3.7 & 2 \\
CD-48 2972 & 250 & 1 &  -3.1 & 2 \\
CPD-54 1295 & 222 & 6 & -3.89 & 6 \\
HD 61005 & 173 & 7,8 & -4.09 & 10 \\ 
CD-48 3199 & 230 & 1 & -3.3 & 2 \\
CD-43 3604 & 320 & 1 & -2.76 & 10 \\
CD-58 2194 & 270 & 1 & -3.17 & 2 \\
CD-57 2315 & 308 & 1 & -3.65 & 2 \\
CPD-62 1197 & 280 & 1 & -3.19 & 2 \\
HD 84075 & 170 & 9  & -4.28 & 10 \\
BD-20 2977 & 260 & 1 & -3.39 & 2 \\
CD-39 5833 & 260 & 1 & -3.10 & 10 \\
HD 85151A & 220 & 1 & -3.71 & 10 \\
HD 85151B & 250 & 1 &  -3.71 & 10 \\
HD 309851 & 170 & 1 & -3.71 & 2,10 \\
TYC 8617-909-1 & 43 & 6 & -3.49 & 6 \\
CD-42 7422 & 260 & 1 & -3.25 & 2 \\
CD-75 652 & 200 & 1 & -3.43 & 2 \\
HD 129496 & 150 & 1 & -3.44 & 10 \\
NY Aps & 182 & 1 & -3.35 & 2 \\
2MASSJ1931-2134 & & & -3.26 & 4 \\
CD-52 9381 & 60 & 1 & -2.93 & 2 \\

\br
\end{tabular}
\end{table}
\noindent Notes to Table 4 $-$ References: (1) Torres et al 2006, (2) Kiraga  2012, (3) Reidel et al 2011, (4) Riaz et al 2006, (5) C. Melis 2017 (private communication), (6) Moor et al 2013, (7) De Silva et al 2013, (8) White et al 2007, (9) Da Silva et al 2009, (10) I. Song 2018 (private communication).

\clearpage

\begin{table}
\caption{Luminosity vs Vsini and Lithium Equivant Width}
\begin{tabular}{@{}lccccc}
\br
Star & Li 6707 $\AA$ & Vsini & Star & Li 6707 $\AA$  & Vsini   \\
high Lum &  EW (m$\AA$) & (km/s)   & low Lum & EW (m$\AA$) & (km/s)  \\
\mr
CD-43 3604 & 320 & 40.4 & CD-29 2360 & 180 &   \\
CD-58 2194 & 270 & 85 & CD-42 2906 & 275 & 10.8 \\
CD-57 2315 & 308 & 24 & CD-48 2972  & 250 & 52 \\
CPD-62 1197 & 280 & 84 & CPD-54 1295 & 222&  \\
HD85151 & 235 & & HD 61005 & 173 & 8.1 \\
CD-42 7422 & 260 & 28 & CD-48 3199 & 230 & 24.7 \\
& & & BD-20 2977 & 260 & 10.1 \\
& & & CD-39 5833 & 260 & 10.5 \\
& & & CD-75 652 & 200 & 20.1 \\
& & & NY Aps & 182 & 10.8 \\

\br
\end{tabular}
\end{table}
\noindent Notes to Table 5 $-$ The ``high Lum'' stars sit well above the 40 Myr isochrone in Figures 2 \& 3.   The ``low Lum'' stars are found near the 40 Myr isochrone. 

\clearpage



\section{Notes on likely Argus members}

If a star in Table 1 does not appear in either Appendix A or B, then we consider it to be an Argus member but with nothing of interest worthy of commentary.

49 Cet:  Excess IR emission.  If indeed a member of Argus, 49 Cet is currently the oldest known main-sequence star (Zuckerman \& Song 2012) that is surrounded by sufficiently abundant molecular gas (CO) to be detectable with a radio telescope (Zuckerman et al 1995).  

CD-29 2360: H$\alpha$ emission (EW = 100 m$\AA$; Torres et al 2006)

AP Col:  H$\alpha$ emission (EW = -6 to -35 $\AA$; Reidel et al 2011)  

TYC 155-2167-1:  H$\alpha$ emission (EW = 675 m$\AA$) is broad and double peaked in an APF spectrum obtained at Lick Observatory on 8 Dec 2017 (UT) by Carl Melis.

CD-42 2906:  6” binary. The G magnitude of the primary is 10.4 and of the secondary 14.3 (Table 2).  SIMBAD gives a spectral type of K1 for the primary, so the secondary is of type early M.  Both stars are plotted in Figures 2 and 3.

CD-48 2972:  Binary with HD 59659

CPD-54 1295: 3.4” binary.  The G magnitude of the primary is 10.3 and of the secondary 15.8.  The color of the primary corresponds to a late-G type star, so the secondary is of type early M.

HD 61005; G8-type.  Excess IR emission.

CPD-62 1197:  the star is incorrectly called CD-62 1197 in Torres et al (2008) and Elliot et al (2016).

HD 84075:  G2-type.  Excess IR emission

HD 88955:  Excess IR emission.  The star plots high on a M$_{V}$ vs B-V color-magnitude diagram (Figure 1 in Zuckerman \& Song 2012) suggesting that it might be older than other A-type stars proposed as members of Argus.  But in Figures 2 \& 3 and also on an absolute G vs G-G$_{RP}$ diagram kindly provided by Jonathan Gagne, the location of the star is consistent with other A-type members of Argus.  The star's V velocity (Table 3) is somewhat less negative than the V of most Argus stars, but with a large error bar.  Overall, we consider HD 88955 to be a probable member of Argus.

HD 90874:  Excess IR emission

TYC 8617-909-1: H$\alpha$ emission (EW = 970 m$\AA$ ; Moor et al 2013)

HD 101615 :  UVW marginal (see Table 3).  Small vsini. May not be an Argus member.

HD 102647 :  $\beta$ Leo.   Excess IR emission.  No Gaia DR2 data.
With Hipparcus data and a radial velocity of -0.2 km/s (Gontcharov 2006), UVW = -20.1, -16.1, -7.8.  
 $\beta$ Leo is included in the calculations of mean UVW and mean distance from Earth of Argus field stars (see Notes to Table 3).  In spite of the lack of Gaia data, it is still possible to position $\beta$ Leo accurately in Figures 2 and 3.  To derive its G magnitude we utilize Argus A-type stars HD 90874 and HD 145689 which bracket $\beta$ Leo in Tycho B$_{t}$ minus V$_{t}$.  The V magnitude of each of these two HD stars is equal to its Gaia DR2 G magnitude.  Therefore, we can safely assume that the G magnitude of  $\beta$ Leo is equal to its V magnitude.  Its V magnitude -- which can be derived from its Tycho B$_{t}$ and V$_{t}$ magnitudes -- is equal to 2.13, with negligible error.  The Ks magnitude of $\beta$ Leo given in the 2MASS catalog is 1.88 +/- 0.19.  But one can do better by going to the Two-Micron Sky Survey catalog (Neugebauer \& Leighton 1969) where the K magnitude is given as 1.90 +/- 0.05.  For an A3-type star the 2-$\mu$m Sky Survey K-magnitude is equal to the Ks magnitude in 2MASS.  So we use Ks = 1.9 to position $\beta$ Leo in Figures 2 and 3.

CD-75 652:  22” binary. Both stars are plotted in Figures 2 \& 3.

HD 123058:  F4-type.  Located in the Galactic plane, thus difficult to determine if the star has excess IR emission.  

HD 129496:  SB1?

HD 138965:  Excess IR emission

HD 145689:  Excess IR emission

HD 168913:  plots high on Figures 2 \& 3.   A7 + A9 according to SIMBAD.  After correction for binarity, the location of HD 168913 in Figures 2 \& 3 is consistent with that of other Argus members. 

HD 188228:  Excess IR emission.

\section{Notes on less likely and unlikely Argus members}

BW Phe: has sometimes been considered as a possible Argus member.  But the UVW (-15, -8, -1) is wrong and the Li EW = 148 m$\AA$ (Torres et al 2006) is characteristic of a star of Pleiades age.

HD 67945: FO type star.  There is a huge difference between parallax and proper motions given in the Gaia DR1 and DR2 catalogs.  With the DR1 values, UVW =  -23.5, -13, -6 (consistent with Argus membership), but with DR2, UVW = -36, -7, -12.  De Silva et al (2013) suggest that the star might be an SB2 although there is no indication of this in other published papers.  HD 67945 is not plotted on Figures 2 \& 3.  With Ks$_{abs}$ = 1.89 and G-Ks = 0.85, as may be seen, HD 67945 would plot a bit higher than the other stars in Figures 2 \& 3.

NLTT 20303:   Shkolnik et al (2012) suggested this star as a possible Argus member.  There is a peculiarity in the Gaia DR2 database:  there is a somewhat fainter star about 8” away with large and similar, but not exact, parallax and proper motion; the difference in the proper motion in declination is far outside of the quoted errors.  With Gaia DR2 values for the “primary” including the RV = 20.5 km/s, then UVW = -24.9, -15.7, -13.7.  If instead one averages the DR2 values for the two stars, then UVW = -27, -15, -13.  In either case, the W velocity is off from that of the Argus Association.  Plotting the star on Figure 2 suggests that it is somewhat older than Argus.

CD-74 673:  this star has been suggested to be an Argus member in various papers.  However, with the Gaia parallax, no combination of proper motions and radial velocities given in the literature can match the Argus UVW.  A typical UVW would be -31, -21, 0.  Nonetheless, the star is obviously young.  The Li EW is 230 m$\AA$ (Torres et al 2006) and Lx/Lbol = -3.04 (Kiraga 2012).  The star plots between the 20 and 40 Myr isochrones in Figure 2.

HD 188728:  The star plots high on a M$_{V}$ vs B-V color-magnitude diagram (Figure 1 in Zuckerman \& Song (2012) suggesting that it might be older than other A-type stars proposed as members of Argus.  It also plots high in Figures 2 \& 3 and on the absolute G vs G-G$_{RP}$ diagram referred to above for HD 88955.  Thus, notwithstanding the reasonable agreement of UVW of HD 188728 with that of the Argus Association (Table 3), we consider HD 188728 to be at best a possible member of Argus.

CD-52 9381: The star plots low -- near 100 Myr -- on Figures 2 and 3. The EW (60m$\AA$) of the Li 6707 $\AA$ line could be consistent with a star of Tuc/Hor age or of Pleiades age (see Figure 3 in Zuckerman $\&$ Song 2004).  The same is true of the X-ray luminosity (Figure 4 in Zuckerman $\&$ Song 2004).  Thus CD-52 9381 might be significantly older than Argus members.

HD 192640:   The star plots high on a M$_{V}$ vs B-V color-magnitude diagram (Figure 1 in Zuckerman \& Song (2012) suggesting that it might be older than other A-type stars proposed as members of Argus.  It also plots high in Figures 2 \& 3 and on the absolute G vs G-G$_{RP}$ diagram referred to above for HD 88955.  Thus, notwithstanding the reasonable agreement of UVW of HD 192640 with that of the Argus Association (Table 3), we consider HD 192640 to be at best a possible member of Argus.

\end{document}